\documentclass[twocolumn, times]{aastex61}
\usepackage{natbib}
\received{xxx}
\revised{xxx}
\accepted{xxx}
\submitjournal{ApJ}

\shorttitle{Multi-frequency VLBA study of 1H\,0323+342}
\shortauthors{Hada et al.}

\begin{document}

\title{Collimation, acceleration and recollimation shock in the jet of gamma-ray emitting radio-loud narrow-line Seyfert 1 galaxy 1H\,0323+342}

\correspondingauthor{Kazuhiro Hada}
\email{kazuhiro.hada@nao.ac.jp}

\author{Kazuhiro Hada}
\affil{Mizusawa VLBI Observatory, National Astronomical Observatory of Japan, Osawa, Mitaka, Tokyo 181-8588, Japan}
\affil{Department of Astronomical Science, The Graduate University for
Advanced Studies (SOKENDAI), 2-21-1 Osawa, Mitaka, Tokyo 181-8588, Japan}

\author{Akihiro Doi}
\affil{Institute of Space and Astronautical Science, Japan Aerospace
Exploration Agency, 3-1-1 Yoshinodai, Chuo, Sagamihara 252-5210, Japan}

\author{Kiyoaki Wajima}
\affil{Korea Astronomy and Space Science Institute (KASI), 776 Daedeokdae-ro, Yuseong-gu, Daejeon 305-348, Republic of Korea}

\author{Filippo D'Ammando}
\affil{INAF Istituto di Radioastronomia, via Gobetti 101, I-40129 Bologna,
Italy}

\author{Monica Orienti}
\affil{INAF Istituto di Radioastronomia, via Gobetti 101, I-40129 Bologna,
Italy}

\author{Marcello Giroletti}
\affil{INAF Istituto di Radioastronomia, via Gobetti 101, I-40129 Bologna,
Italy}

\author{Gabriele Giovannini}
\affil{INAF Istituto di Radioastronomia, via Gobetti 101, I-40129 Bologna,
Italy}
\affil{Dipartimento di Fisica e Astronomia, Universit\`a di Bologna, via
Ranzani 1, I-40127 Bologna, Italy}

\author{Masanori Nakamura}
\affil{Institute of Astronomy \& Astrophysics, Academia Sinica, 11F of Astronomy-Mathematics Building, AS/NTU No. 1, Taipei 10617, Taiwan}

\author{Keiichi Asada}
\affil{Institute of Astronomy \& Astrophysics, Academia Sinica, 11F of Astronomy-Mathematics Building, AS/NTU No. 1, Taipei 10617, Taiwan}

\begin{abstract}
We investigated the detailed radio structure of the jet of 1H\,0323+342 using high-resolution multi-frequency Very Long Baseline Array observations. This source is known as the nearest $\gamma$-ray emitting radio-loud narrow line Seyfert 1 (NLS1) galaxy. We discovered that the morphology of the inner jet is well characterized by a parabolic shape, indicating the jet being continuously collimated near the jet base. On the other hand, we found that the jet expands more rapidly at larger scales, resulting in a conical-like shape. The location of the ``collimation break'' is coincident with a bright quasi-stationary feature at 7\,mas from core (corresponding to a deprojected distance of the order of $\sim$100\,pc), where the jet width locally contracts together with highly polarized signals, suggesting a recollimation shock. We found that the collimation region is coincident with the region where the jet speed gradually accelerates, suggesting the coexistence of the jet acceleration and collimation zone, ending up with the recollimation shock, which could be a potential site of high-energy $\gamma$-ray flares detected by the $Fermi$-LAT. Remarkably, these observational features of the 1H\,0323+342 jet are overall very similar to those of the nearby radio galaxy M87 and HST-1 as well as some blazars, suggesting that a common jet formation mechanism might be at work. Based on the similarity of the jet profile between the two sources, we also briefly discuss the mass of the central black hole of 1H\,0323+342, which is also still highly controversial on this source and NLS1s in general. 
 
\end{abstract}

\keywords{galaxies: active --- galaxies: Seyfert --- galaxies: jets --- radio
continuum: galaxies --- galaxies: individual (1H\,0323+342) --- gamma rays: galaxies}



\section{Introduction} \label{sec:intro}

Narrow-line Seyfert 1 (NLS1) galaxies are a subclass of active galactic nuclei
(AGN) that is defined by its optical properties, i.e., narrow permitted lines
FWHM (${\rm H}\beta <2000$\,km\,s$^{-1}$), [\ion{O}{3}]/H$\beta < 3$, and a
bump due to \ion{Fe}{2} emission lines~\citep{osterbrock1985}. Based on
these peculiar optical characteristics along with the large soft X-ray excess~\citep[e.g.,][]{boller1996}, NLS1s are suggested to have small black hole masses ($10^{6-8}\,M_{\odot}$) and high accretion rates (close to or above the Eddington limit), making this population a possible candidate for studying the physics of rapidly evolving supermassive black holes (SMBHs).  

In recent years, this class of objects attracts a great deal of attention in the
high-energy astrophysical community; the discovery of $\gamma$-ray emission by the $Fermi$ satellite from some radio-loud NLS1s suggested the presence of a possible third class of AGN with relativistic jets, in addition to blazars and radio galaxies~\citep{abdo2009a, abdo2009b}. The $\gamma$-ray emission from NLS1s is surprising and poses a challenge to our current understanding of jet formation, since NLS1s are generally thought to be hosted by spiral galaxies, which are not expected to produce high-power jets. To date, $\gamma$-ray emission has been detected in a dozen of radio-loud NLS1s~\citep{dammando2016, paliya2018}, and the ejection of superluminal features is detected in a few of them~\citep{dammando2013, lister2016}. This further confirms that relativistic jets constitute an essential channel of energy release at least in these active NLS1 nuclei. 

Some pioneering VLBI studies on radio-loud NLS1s~\citep{doi2006, doi2007, doi2011, giroletti2011, foschini2011, orienti2012, orienti2015, lister2016} revealed that the pc-scale morphologies of radio-loud NLS1s are reminiscent of blazars, i.e., consisting of a bright compact core and one-sided jet. This implies that similar processes may be at work in producing $\gamma$-rays between blazars and $\gamma$-ray detected NLS1s, such as the Doppler boosting effect with small jet viewing angles. However, due to the lack of sufficient VLBI studies so far, detailed pc-scale structures of the NLS1s jets are still poorly understood. Direct constraints on the innermost jet properties from VLBI are complementary to what can be inferred from radio-to-$\gamma$-ray broadband SED modeling~\citep{abdo2009b, foschini2011, dammando2012, dammando2013, dammando2015a, dammando2015b, dammando2016}. Moreover, dedicated high-resolution studies of NLS1 jets may help understand the mechanisms of relativistic jet launching, collimation and acceleration in a possible distinct ($M_{\rm BH}$, $\dot{M}$) domain from blazars/radio galaxies which have been intensively investigated~\citep[e.g.,][]{marscher2008, asada2012, hada2013, nagai2014, boccardi2016, gomez2016, mertens2016, nakahara2018, giovannini2018}.

In this paper we focus on a famous NLS1 galaxy 1H 0323+342. This source is known as the nearest (a redshift of 0.063) $\gamma$-ray detected NLS1~\citep{abdo2009b}, and is one of the few NLS1s where the host galaxy is imaged~\citep{zhao2007} along with FBQS~J1644+2619~\citep{dammando2017} and PKS\,2004--447~\citep{kotilainen2017}. Thus 1H\,0323+342 is a unique target that allows us to probe the vicinity of the central engine at the highest linear resolution among $\gamma$-ray active NLS1s (i.e., 1\,mas = 1.2\,pc). The pc-scale structure of the source was first examined by \citet{wajima2014} with the Japanese VLBI Network and the Very-Long-Baseline-Array (VLBA) at 2, 8 and 15\,GHz, revealing a blazar-like one-sided core-jet structure toward the southeast direction. Kinematic properties of the pc-scale jet were investigated by several authors~\citep{wajima2014, fuhrmann2016, lister2016, doi2018}. The latter three papers reported highly superluminal features up to $\sim$9\,$c$. In addition, \citet{wajima2014}, \citet{fuhrmann2016} and \citet{doi2018} found one or two quasi-stationary features at parsec scales that might be associated with standing shocks. Detailed radio-to-$\gamma$-ray broadband properties of the source are studied by \citet{paliya2014}, \citet{yao2015} and \citet{kynoch2018}. Regarding the mass of the central BH, several independent measurements suggest $M_{\rm BH}\sim 2\times 10^{7}\,M_{\odot}$~\citep{zhao2007, yao2015, landt2017, wang2016}, although this is still under active debate and some authors suggest a much larger black hole mass~\citep[a few times of $10^{8}\,M_{\odot}$;][]{tavares2014}.  

In this paper, we report the result from our multi-frequency VLBA observation of 1H\,0323+342, which aims to better understand the pc-to-subpc-scale structures of this jet. In the next section we describe our observation and data reduction. In Sections 3 and 4, our results are presented and discussed. In the final section we will summarize the paper. Throughout the paper we adopt cosmological parameters $H_{\rm 0} = 70$\,km\,s$^{-1}$\,Mpc$^{-1}$, $\Omega_{\rm M} = 0.3$ and $\Omega_{\rm \Lambda} = 0.7$. This corresponds to 1\,mas = 1.2\,pc at $z$ = 0.063. Spectral index $\alpha$ is defined as $S_{\nu}\propto \nu^{+\alpha}$.

\section{Observations and Data Reduction} \label{sec:obs}
We observed 1H\,0323+342 with VLBA on January 2nd 2016. The observations were made at nine frequencies of 1.4/1.7\,GHz (L-band), 2.3/8.4\,GHz (S/X bands), 5\,GHz (C-band), 12/15\,GHz (U-band), 24\,GHz (K-band) and 43\,GHz (Q-band) in a 11-hr continuous run. The receiver at each frequency band was alternated in turn every 10--30 minutes. The observation involved rapid switching between the target and the nearby radio source 0329+3510 (separated by 1.36\,deg on the sky) to allow phase-referencing and relative astrometry between the two sources. Some short scans on the bright source 3C\,84 were also inserted as a fringe finder and bandpass calibrator. The observation was made in dual (left/right-hand circular) polarization mode. The received signals were sampled with 2-bit quantization and recorded at an aggregate rate of 2\,Gbps (a total bandwidth of 512\,MHz) using the digital-down-converter-4 (DDC-4) wideband recording mode. The down-converted signals were divided into four 64 MHz sub-bands in each polarization. 

The initial data calibration was performed using the National Radio Astronomy Observatory Astronomical Image Processing System (AIPS) based on the standard VLBI data reduction procedures. The amplitude calibration was made with opacity corrections using the measured system noise temperature and the elevation-gain curve of each antenna. We next performed a priori corrections for the visibility phases; antenna parallactic angle differences between 1H\,0323+342 and 0329+3510, ionospheric dispersive delays using the ionospheric model provided by the Jet Propulsion Laboratory, and instrumental delays/phases using a scan of 3C\,84 were corrected. Then, to create images of these sources, we performed a fringe-fitting on each source separately and removed residual delays, rates, and phases assuming a point source model. Images were created in the DIFMAP software~\citep{shepherd1997} with iterative phase/amplitude self-calibration. 

In this paper, we primarily focus total-intensity results. Our dedicated multi-frequency polarimetric study based on these data will be presented in a forthcoming paper.

\begin{table}[ttt]
 \begin{minipage}[t]{1.0\columnwidth}
  \centering 
  \caption{VLBA observations of 1H\,0323+342 on 2016 January}
    \begin{tabular*}{1.0\columnwidth}{@{\extracolsep{\fill}}ccccc}
    \hline
    \hline
     Frequency & Beam size & $I_{\rm p}$    & $I_{\rm
     rms}$ \\
         (GHz) & (mas$\times$mas, deg.) & (mJy/bm) & (mJy/bm) \\
               & (a)     & (b)     & (c)    \\
    \hline
     1.44   & 9.92 $\times$ 5.69, $0$ & 135  & 0.085 \\
     1.70   & 8.58 $\times$ 4.84, $-4$ & 124  & 0.134 \\
     2.32   & 6.80 $\times$ 3.61, $0$ & 146  & 0.193 \\
     5.00   & 3.09 $\times$ 1.64, $-8$ & 155  & 0.056 \\
     8.68   & 1.78 $\times$ 0.94, $-2$ & 177  & 0.141 \\
     12.5   & 1.14 $\times$ 0.64, $-4$ & 277  & 0.118 \\
     15.5   & 0.96 $\times$ 0.53, $-6$ & 283  & 0.144 \\
     23.9   & 0.61 $\times$ 0.32, $-12$ & 280  & 0.224  \\
     43.3   & 0.36 $\times$ 0.18, $-5$ & 242  & 0.318  \\
    \hline
    \end{tabular*} \medskip
  \end{minipage}
  \label{tab:img_prm} Notes: (a) synthesized beam with naturally-weighted scheme: (b) peak intensity of self-calibrated images of 1H\,0323+342 under naturally-weighting scheme: (c) rms image noise level of 1H\,0323+342 images under naturally-weighting scheme.
\end{table}

\begin{figure*}[ttt]
 \centering
 \includegraphics[angle=0,width=1.0\textwidth]{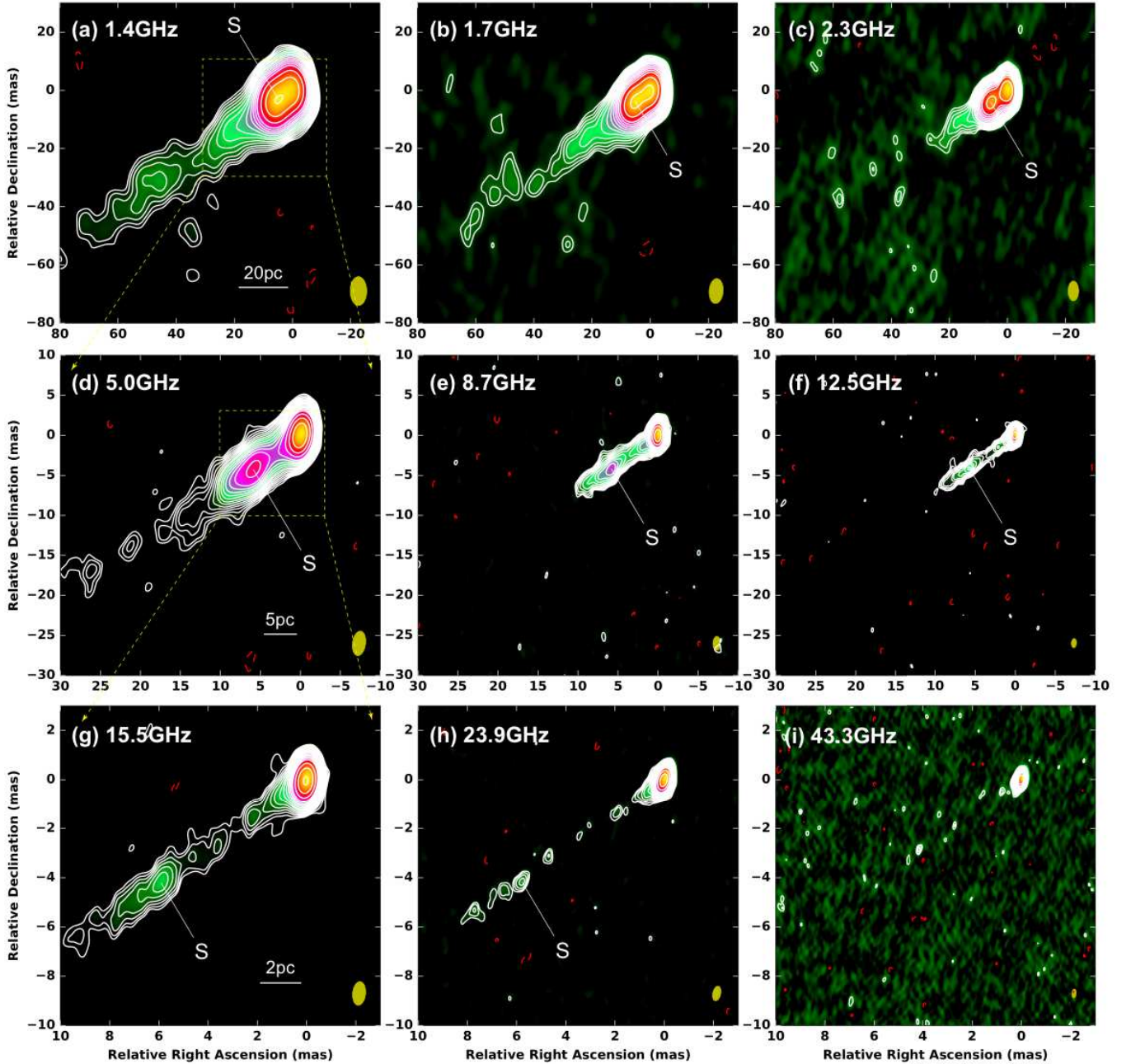}
 \caption{VLBA images of 1H\,0323+342. (a),(b),(c),(d),(e),(f),(g),(h) and (i) are naturally-weighting images at 1.4, 1.7, 2.3, 5.0, 8.7, 12.5, 15.5, 23.9, 43.3\,GHz, respectively. For each image, contours start from -1, 1, 2,...times a 3$\sigma$ image rms level and increase by factors of $2^{1/2}$. The yellow ellipse at the bottom-right corner of each map is the beam size of the map. The feature S is a quasi-stationary component located at 7\,mas from the core.}
 \label{fig:vlbaimages}
\end{figure*}

\section{Results}

\subsection{Multi-frequency images}
In Figure\,\ref{fig:vlbaimages} we show VLBA images of 1H\,0323+342 obtained at the nine frequencies. At all the frequencies a core-jet structure is seen and the one-sided jet is elongated toward the south-east direction (P.A. $\sim$125\,deg). The overall jet morphology at 8/12/15\,GHz is very similar to the previous VLBA 8/15\,GHz images presented by \citet{wajima2014},  \citet{fuhrmann2016} and \citet{doi2018}. \citet{wajima2014} and \citet{doi2018} reported a quasi-stationary bright feature at $\sim$7\,mas downstream of the core. The corresponding feature consistently exists also in our 8/12/15\,GHz images, indicating that the feature is still present after more than 5--10 years. In the present paper we call this feature as ``S''. 

At low frequencies the feature S is more prominent, becoming the brightest component in the 1.4\,GHz image. The low frequency images also reveal the downstream jet emission extending $\sim$60-80\,mas from the core. In contrast, the source structure at 24 and 43\,GHz becomes more compact and mostly core-dominated. 

We stress that this is the first VLBI result on 1H 0323+342 up to 43\,GHz where the highest resolution and transparency to the jet base is attained. To quantify the innermost radio core at 43\,GHz, we fit the calibrated visibility data with an elliptical Gaussian model using the DIFMAP task {\tt modelfit}. Our best-fit result is summarized in Table~\ref{tab:43Gcore}. By using these values and the equation (1) in \citet{wajima2014}, the brightness temperature of the 43\,GHz core is estimated to be $T_{\rm B} = 5.3 \times 10^{10}$\,K. The value is very similar to those reported by \citet{wajima2014} at low frequencies.

Note that \citet{wajima2014} and \citet{fuhrmann2016} reported another quasi-stationary feature around 0.3-0.5\,mas from the core in their multi-epoch 8/15\,GHz images. Here we searched for the corresponding feature in our images at 8/12/15/24/43\,GHz. However, the jet intensity profile around 0.3-0.5\,mas was more or less smooth in our images and we did not find such a feature clearly.

\begin{table}[ttt]
 \begin{minipage}[t]{1.0\columnwidth}
  \centering \caption{Modelfit parameters for 43\,GHz core}
    \begin{tabular*}{1.0\columnwidth}{@{\extracolsep{\fill}}cccccc}
    \hline
    \hline
    $\theta_{\rm maj}$  & $\theta_{\rm min}$ & PA & $S_{\rm core}$  \\
    ($\mu$as) & ($\mu$as) & (deg.) & (mJy)     \\
    (a)  &   (b)  & (c)  & (d)      \\
    \hline
    $131\pm9$ & $60\pm10$ & $130\pm3$ & $294\pm29$   \\
    \hline
    \end{tabular*}
  \end{minipage}
  \label{tab:43Gcore} Notes: (a) major axis FWHM size of the derived elliptical
  Gaussian; (b) minor axis FWHM size of the derived elliptical Gaussian; (c) position angles of the major axes of the Gaussian model; (d) total flux densities of the Gaussian models; (e) combined visibility data over the two epochs. For (a), (b) and (c), a practical uncertainty of each parameter was estimated by comparing the difference of the derived results between modelfit in Difmap and JMFIT in AIPS. For (d), we adopt 10\% uncertainty based on the typical amplitude calibration accuracy. 
\end{table}

\begin{figure*}[ttt]
 \centering
 \includegraphics[angle=0,width=0.8\textwidth]{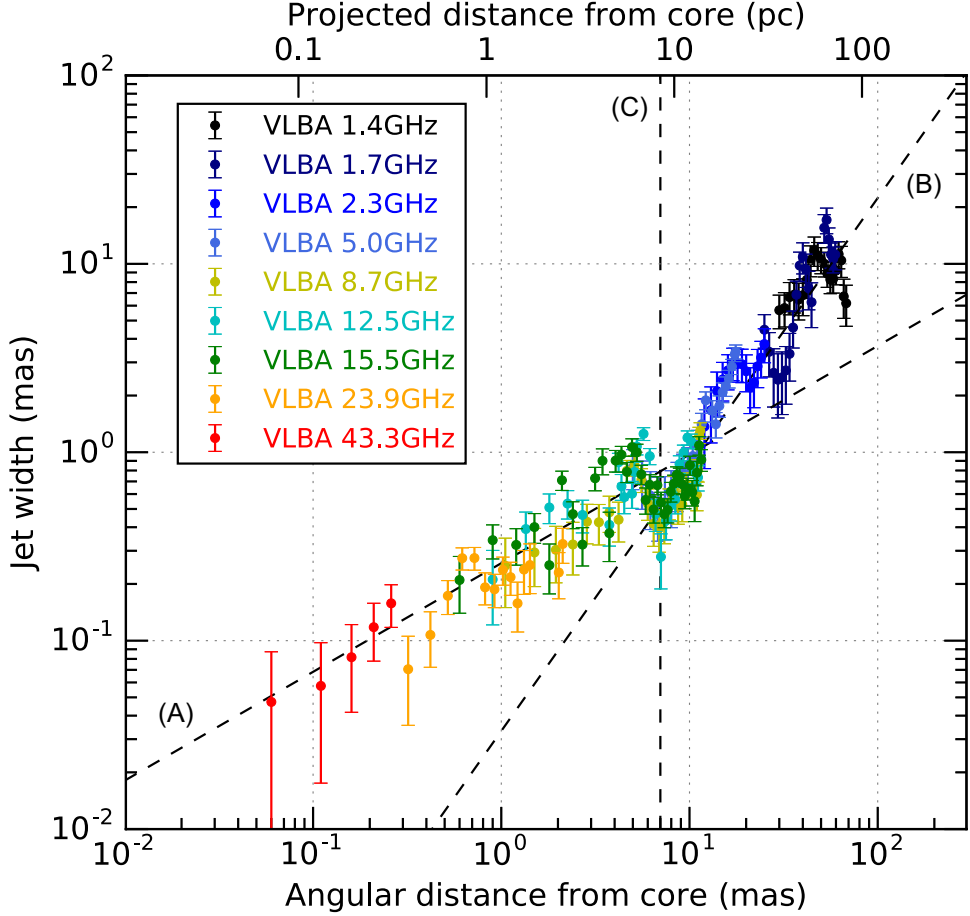}
 \caption{Jet width profile $W(z)$ of 1H\,0323+342 as a function of (projected) distance $z$ from the 43\,GHz core. The dashed line (A) represents a best-fit model for the inner ($z<6.5$\,mas) jet $W(z)\propto z^{0.58}$, indicating a parabolic collimating jet. The dashed line (B) represents a best-fit model for the outer ($z>7.5$\,mas) jet $W(z)\propto z^{1.41}$, indicating a hyperbolic expanding jet. The vertical dashed line (C) indicates the location of the quasi-stationary feature S, where the jet width is locally shrinking (by a factor of 2) and the transition of jet shape occurs.}
 \label{fig:collimation}
\end{figure*}

\subsection{Jet width profile}
Measuring the profile of jet width $W(z)$ as a function of radial distance $z$ provides important clues to the formation and collimation mechanisms of a relativistic jet, as demonstrated in recent VLBI studies on nearby radio galaxies~\citep{asada2012, hada2013, nakamura2013, hada2016, boccardi2016, tseng2016, nakahara2018, giovannini2018}. Our multi-frequency images on 1H\,0323+342 allow us to examine $W(z)$ on this jet in detail. 

We basically followed the methodology in our previous studies on M87~\citep{hada2013, hada2016}. We made transverse slices of the 1H\,0323+342 jet every 1/2--1/3 of the beam size along the jet and each slice was fit with a single Gaussian function (because the jet is single-ridged at all distances in our VLBA images\footnote{Note that \citet{doi2018} has recently revealed a limb-brightened jet profile around S based on a detailed image analysis of VLBA archival data. This does not affect the result of our jet-width measurement presented here.}). We then defined the jet width at each $z$ as a deconvolved size of the fitted Gaussian $(\Theta_{\rm Gauss}^2-\Theta_{\rm beam}^2)^{1/2}$ where $\Theta_{\rm Gauss}$ and $\Theta_{\rm beam}$ are FWHM of the fitted Gaussian and the projected beam size perpendicular to the jet, respectively. The result is summarized in Figure~\ref{fig:collimation}, where the measured jet width profile is plotted as a function of projected (angular) distance from the innermost (43\,GHz) core. Thanks to the wide-frequency coverage, we could measure the jet width profile over $\sim$3 orders of magnitude in distance from $\lesssim$$0.1$\,mas to $\sim$80\,mas. 

Figure~\ref{fig:collimation} revealed some very interesting features on the structure of this jet. The jet shows a clear change in the collimation slope between $z\lesssim7$\,mas and $z\gtrsim7$\,mas. We then fit a power-law function $W(z) = A z^{a}$ to the inner jet data ($0.1\,{\rm mas}<z<6.5\,{\rm mas}$) and the outer jet data ($7.5\,{\rm mas}<z<80\,{\rm mas}$) separately (where $A$ and $a$ are free parameters). We found that the inner jet region within 7\,mas is well described by a parabolic shape with $a=0.58\pm0.02$, indicating the jet being continuously collimated over these distances. On the other hand, beyond 7\,mas the jet starts to expand more rapidly with $a=1.41\pm0.02$, suggesting a conical or hyperbolic expansion. This parabolic-to-conical/hyperbolic transition occurs exactly at 7\,mas from the core, where the bright stationary component S is located. The jet width at 7\,mas (on the stationary component) is locally shrinking by a factor of 2. The local contraction at S has recently been reported by \citet{doi2018} based on MOJAVE 15\,GHz data, and the present result further confirms this feature. All of the findings reported here are robust thanks to the sufficient number of dataset and consistency at different frequencies. It is notable that the observed jet collimation profile looks very similar to that seen in the nearby radio galaxy M87 and HST-1~\citep{asada2012}. We will discuss the comparison between M87 and 1H\,0323+342 in Section~4. 

\begin{figure}[ttt]
 \centering
 \includegraphics[angle=0,width=1.0\columnwidth]{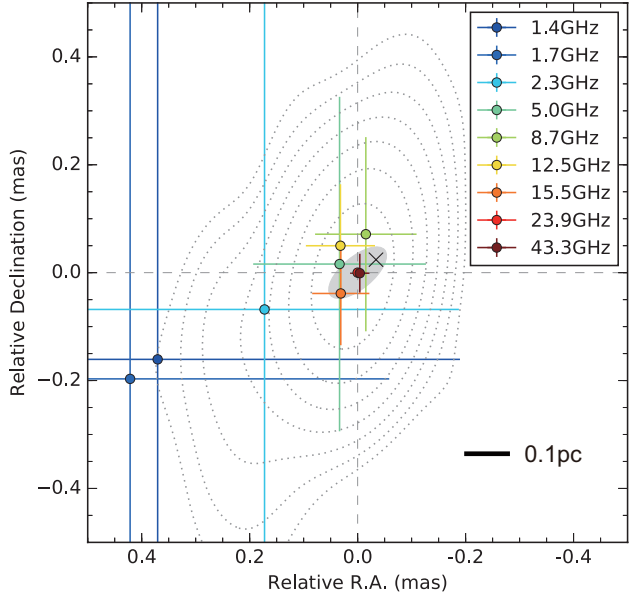}
 \caption{Astrometry of the core position of 1H\,0323+342 at various frequencies. Position of the 24\,GHz core is set to be the coordinate origin of this plot. Note that position of the 43\,GHz core is almost identical to that of the 22\,GHz core. The contours are a total intensity map 1H\,0323+342 at 43\,GHz. Grey-shaded ellipse indicates a best-fit elliptical Gaussian model to the 43\,GHz core (see Table~\ref{tab:43Gcore}). The cross symbol indicates an extrapolated position $z_0$ of the origin of the parabolic jet (see Section~3.3), which is a likely position of the central black hole.}
 \label{fig:0321corepos}
\end{figure}

\subsection{Core-shift}
Our multi-frequency VLBA data also enable us to examine the core-shift~\citep[e.g.,][]{lobanov1998, hada2011} of 1H\,0323+342. We investigated the core-shift of the source in the following way. Between 24 and 43\,GHz, the core-shift of the target was measured with respect to the calibrator 0329+3510\footnote{In this case, the core-shift from the calibrator may affect the measurement of the core-shift of the target, but we confirmed that the calibrator's core-shift between 24 and 43\,GHz is negligible (see Appendix for details)} by using the phase-referenced images of 0329+3510. On the other hand, between 1.4 and 24\,GHz, the complex structure of the calibrator prevented us from reliable core-shift measurements (see Appendix). Instead, we performed a 'self-referencing' method at these frequencies because 1H\,0323+342 shows a well-defined optically-thin feature S at $\sim$7\,mas from the core. We thus measured the core-shift between 1.4 and 24\,GHz by assuming that the feature S has no frequency-dependent position shifts. 

The result of our core-shift measurement is shown in Figure\,\ref{fig:0321corepos}. The origin of this plot is set to be the core position at 24\,GHz, at which core-shift measurements could be made with respect to all the other frequencies. The $(x, y)$ position errors for each data are one-tenth of the beam size, which is rather conservative. Interestingly, the measured core positions between 5.0 and 43\,GHz were concentrated within the central $\sim$100\,$\mu$as area and we did not find any clear trend of core-shift among these frequencies. At 2.3\,GHz and the lower frequencies, Figure\,\ref{fig:0321corepos} implies a possible trend of core-shift in the direction of the jet, but the position uncertainties of the low-frequency cores are quite large due to the poor angular resolution and the severe blending between the core and neighboring jet emission. Therefore, we conclude that the core-shift $\Delta \theta_{\rm cs}$ of 1H\,0323+342 is very small: $\Delta \theta_{\rm cs} \lesssim 100$\,$\mu$as between 5.0 and 43\,GHz and $\Delta \theta_{\rm cs}  \lesssim 500$\,$\mu$as between 1.4 and 43\,GHz. 

Given that the radio core shows an optically-thick spectrum (see Section~3.4), the small core-shift suggests the jet having a small viewing angle toward our line-of-sight. This is consistent with the detection of highly superluminal components up to $\sim$9\,$c$~\citep{lister2016} which constrains $\theta_{\rm view}$$\lesssim$12\,deg. An even smaller viewing angle $\theta_{\rm view}$$\lesssim$4\,deg is suggested if the time scales of radio variability are additionally considered~\citep{fuhrmann2016}.

\begin{figure}[ttt]
 \centering
 \includegraphics[angle=0,width=1.02\columnwidth]{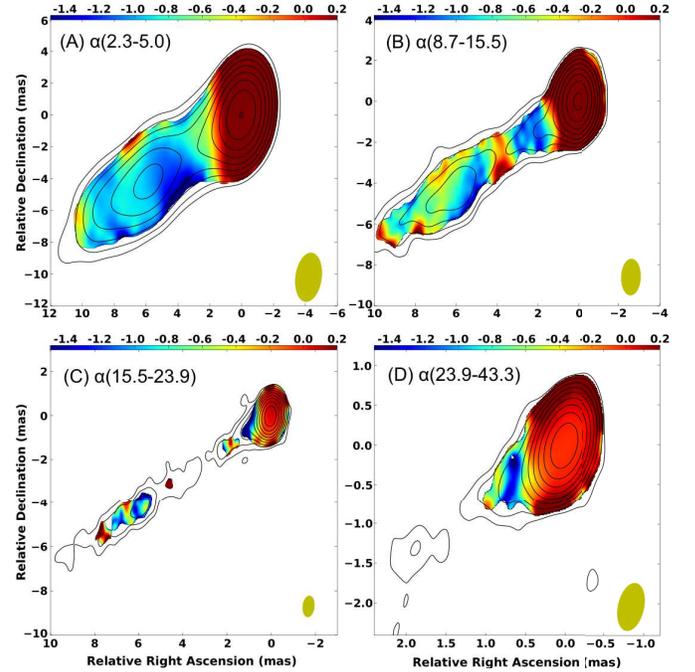}
 \caption{Spectral index (defined as $S_{\nu}\propto\nu^{+\alpha}$) distribution of 1H\,0323+342 for various frequency pairs (color maps): (A) between 2.3 and 5.0\,GHz; (B) between 8.7 and 15.5\,GHz; (C) 15.5 and 23.9\,GHz; (D) 23.9 and 43.3\,GHz. Contours in each inset are a total intensity map at the lower frequency, starting from (1, 2, 4, 8, 16, 32, 64,...) times 0.2\,mJy/beam. Yellow ellipse in each inset is the size of convolving beam.}
 \label{fig:spectra}
\end{figure}

The small core-shift and viewing angle implies that the angular separation of the central BH with respect to the radio core is rather short. Such a separation can be measured if the frequency dependence of the core-shift is accurately determined~\citep{hada2011}, while the accuracy of our core-shift measurements on 1H\,0323+342 is inadequate to use this method. Instead, we can reasonably deduce the BH position by taking advantage of the observed parabolic profile of the inner jet. Assuming that the parabolic shape continues all the way down to the BH and the jet width converges to zero at the BH position, we again fit the inner jet within $z<6.5\,{\rm mas}$ using the following equation (instead of $W = Az^a$): 
\begin{equation}
W(z) = A (z-z_0)^a, 
\end{equation}
where the additional free parameter $z_0$ is the angular distance of the BH position from the 43\,GHz core. We obtained a very similar parabolic profile $a=0.60\pm0.03$ together with $z_0=-41\pm36$\,$\mu$as. As visualized in Figure~\ref{fig:0321corepos}, the derived BH position $z_0$ (as shown by a cross symbol) is within the ellipse of the 43\,GHz core. Thus the parabolic jet shape consistently suggests a small viewing angle of this jet. For $\theta_{\rm view}=4-12$\,deg, the deprojected distance between the 43\,GHz core and the BH position is estimated to be $\sim$(0.24--0.71)\,pc. Note that $z_0$ might be an upper limit of the BH-core distance if the (parabolic) jet has a non-zero width at the BH position.

\subsection{Spectra}
In Figure~\ref{fig:spectra}, we show maps of spectral index distribution of 1H\,0323+342 for various frequency pairs. To create an accurate spectral index map for each frequency pair, map positions at the two frequencies were adjusted by using the astrometric results in Section~3.3. Also in Figure~\ref{fig:spectra}, we used the same uv range and pixel sampling for each frequency pair. Then the two maps were convolved with a synthesized beam at the lower frequency. 

As seen in Figure~\ref{fig:spectra}, the core regions generally show flat-to-inverted (i.e., optically-thick) spectra, while the extended jet regions show steep (optically-thin) spectra, which is consistent with typical characteristics of blazar jets. Some previous studies suggest that radio spectra of NLS1 jets are similar to those in young radio galaxies~\citep[e.g.,][]{gallo2006, berton2016b}, whose spectral turnover occurs around 1--10\,GHz. However, our result indicates that the radio spectra of the 1H\,0323+342 core remain flat even at 43\,GHz.  

Regarding the extended jet, there is a possible trend that the spectra on and right down the quasi-stationary feature S are slightly flatter than those in the upstream side of S. Although the uncertainties are still large, this might indicate re-energization of nonthermal electrons via shock in the feature S. 

\section{Discussion}

\subsection{Jet Collimation Break at the Recollimation Shock}
Structural transition of jet profile has recently been discovered in some nearby radio galaxies (M87; \citet{asada2012}, NGC\,6251; \citet{tseng2016}, 3C\,273; \citet{akiyama2018}, NGC\,4261; \citet{nakahara2018}). This has triggered active discussion about the mechanisms of jet formation and collimation. Our result presented here is the first example where such a structural transition was discovered in the jet of a NLS1 galaxy.

\begin{figure}[ttt]
 \centering
 \includegraphics[angle=0,width=1.0\columnwidth]{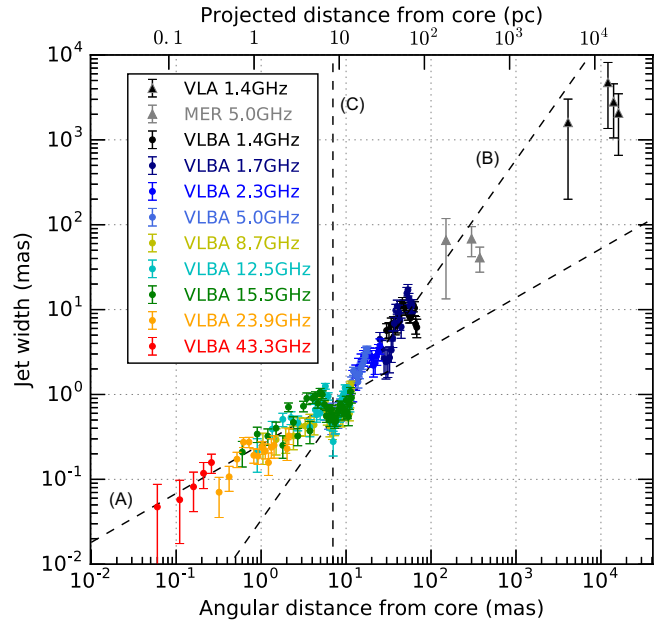}
 \caption{Extended plot of jet width profile of 1H\,0323+342 including MERLIN and VLA measurements at large scales. The dashed lines (A), (B) and (C) are the same as in Figure~\ref{fig:collimation}.}
 \label{fig:collimation_large}
\end{figure}

To check the jet collimation profile beyond the VLBI scales, we additionally analyzed a few archival data of 1H\,0323+342 obtained by connected radio interferometers. We found one VLA data at 1.4\,GHz (Project ID AM601, obtained in 1998 with the B configuration and an on-source time of 15 minutes) and another data with MERLIN at 5\,GHz (obtained in 2006 with an on-source time of 5 minutes). At the large scales, the main jet can continuously be seen up to $\sim$10--20 arcsec from the core, while the overall source morphology shows a hint of the counter jet~\citep[see also][]{anton2008}\footnote{A more detailed study on the kpc-scale structure of this jet will be presented in a forthcoming paper (Doi et al. in preparation)}. In Figure\,\ref{fig:collimation_large} we show an extended plot of the jet width profile (for the main jet) including the VLA/MERLIN measurements. Beyond 100\,mas from the core, the deviation from the inner jet parabola is more evident, and the jet width becomes a factor of $\gtrsim$10 wider than that expected from the parabolic profile.\footnote{There is a possibility that the jet viewing angle is different between pc and kpc scales since the position angle of the kpc-scale jet is $\sim$90\,deg~\citep{anton2008} while the pc-scale jet is $\sim$125\,deg. This might affect the deprojected distance of jet. However, the jet position angle at the MERLIN scales is still very similar to that in the VLBA scales, so the jet profile between VLBA and MERLIN scales should be compared in the same viewing angle.} If we fit the width profile of the outer jet ($z>7.5$\,mas) together with the VLBA/MERLIN/VLA data, a best-fit slope results in $a=1.25\pm0.02$, approaching a conical shape.

The observed jet width profile strongly suggests that the structural transition from parabolic to hyperbolic/conical originates in the quasi-stationary feature S. This situation is remarkably similar to the case of the jet in M87~\citep{asada2012}, where its parabolic to conical transition occurs at the bright quasi-stationary (and possibly $\gamma$-ray emitting) knot HST-1 located at $\sim$120\,pc from the nucleus. For HST-1, its cross section seems to be slightly smaller than that expected from the parabolic profile. This leads to an argument that HST-1 represents a recollimation shock at the end of the collimation zone~\citep{cheung2007, asada2012}. 

For 1H\,0323+342, the local contraction of jet width at the transition site is more evident than HST-1. This feature was first reported by \citet{doi2018} and the present multi-frequency study further confirmed this. While in M87 there is still no clear evidence for the ``recollimation zone'' between the parabolic region and HST-1, in 1H\,0323+342 we can clearly see the recollimation zone just upstream of S ($z$$\sim$6-7\,mas), connecting the whole sequence of parabolic zone, recollimation zone, shock and expansion. The rebrightening of jet intensity with a possible spectral flattening at S indicates the re-energization of jet particles via shock compression. Interestingly, the deprojected distance of the feature S is $\sim$40-120\,pc (for $\theta_{\rm view}\sim 4-12$\,deg), which is similar to that of HST-1 in pc unit. Moreover, S seems to be caused by the passage of highly superluminal components~\citep[][see also Section~4.2]{doi2018}, and this kind of event is  repeatedly seen also in HST-1~\citep{biretta1999, cheung2007, giroletti2012, hada2015}. We therefore suggest that the feature S in 1H\,0323+342 is analogous to HST-1 in M87, and might be a potential site of $\gamma$-ray emission (see Section~4.3).

\begin{figure}[ttt]
 \centering
 \includegraphics[angle=0,width=1.0\columnwidth]{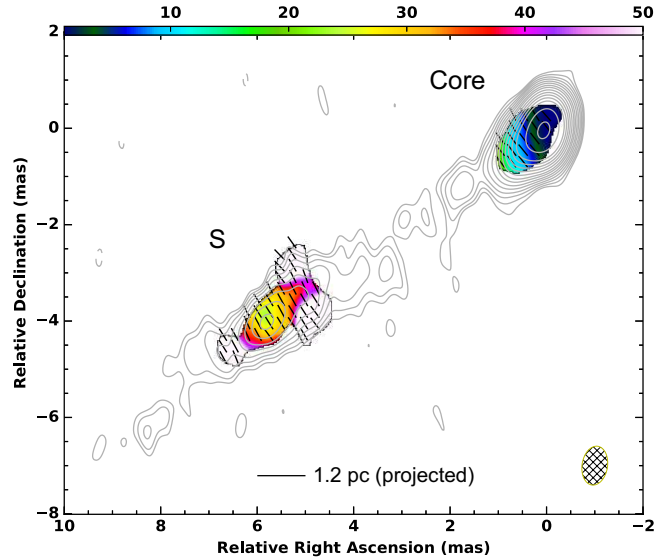}
 \caption{VLBA 15\,GHz polarization map of 1H\,0323+342. We retrieved the calibrated data from the MOJAVE 15\,GHz archive. The total intensity distribution is shown by contours, while the fractional polarization distribution is shown by color (the values are in percent). Note that the EVPA shown are not RM-corrected.}
 \label{fig:pol}
\end{figure}

We further checked one recent VLBA polarization map taken by the MOJAVE 15\,GHz program to search for any polarimetric feature associated with S (Figure~\ref{fig:pol}). Interestingly, the jet exhibits strongly polarized emission exactly on S, where the values of fractional polarization reach as high as $\sim$30\%, indicating the presence of highly ordered magnetic fields. This is in agreement with S being a shocked compressed feature. Note that the apparent electric-vector-polarization angles (EVPA) in both the core and S are perpendicular to the jet, but the EVPA shown are not rotation-measure (RM)-corrected ones, preventing us from discussing the intrinsic geometry of the $B$-fields associated with the jet. A recent optical polarimetry study on this source suggests the optical EVPA being parallel to the jet axis~\citep{itoh2014}, implying the $B$-field direction (associated with the optical emission) being perpendicular to the jet. Thus a future dedicated VLBI polarimetric study including RM measurements would be useful to better understand the $B$-field properties associated with S.

We also note that the presence of a recollimation shock at some distances from the radio core is also suggested in some other blazars and blazar-like radio galaxies such as 3C\,120~\citep{agudo2012}, 3C\,111~\citep{beuchert2018} and CTA\,102~\citep{fromm2011, fromm2013}. In particular, the latter two sources show peculiar time evolution in radio spectra (CTA\,102) or in polarized emission~(3C\,111) when superluminal components pass through a standing component, interpreted as the interaction of moving shocks with a recollimation shock. Application of a similar multi-epoch analysis to 1H\,0323+342 would help confirm the recollimation shock scenario of S and its similarity to blazars.

\subsection{Coexistence of Jet Collimation and Acceleration}

One of the longstanding central questions regarding relativistic jets is the physical link between collimation and acceleration of the flow. This is vital to understand the ultimate mechanisms of jet formation~\citep[e.g.,][]{mckinney2006, komissarov2009} and also the origin/location of $\gamma$-ray emission~\citep[e.g.,][]{marscher2008}. In recent years this topic has extensively been examined for the M87 jet~\citep{nakamura2013, asada2014, mertens2016, hada2017}, and these studies are beginning to uncover the physical link and coexistence of the acceleration and collimation zone (ACZ) in this jet, which is in agreement with magnetic collimation and acceleration models~\citep[e.g.,][]{nakamura2018}. However, hitherto very few observational evidence of ACZ is obtained for other sources.

For 1H\,0323+342, \citet{fuhrmann2016}, \citet{lister2016} and \citet{doi2018} detected significant superluminal motions at parsec scales. Interestingly, they consistently found a trend of gradual increase of jet proper motion with distance. Also, the lower brightness temperature of the core than those of typical $\gamma$-ray blazars indicates a relatively mild Doppler factor in the core region, suggesting that the innermost jet region is still not fully accelerated~\citep{wajima2014, fuhrmann2016}. Therefore, it would be of great interest to compare those velocity results and the collimation profile obtained in our study. 

\begin{figure}[ttt]
 \centering
 \includegraphics[angle=0,width=1.0\columnwidth]{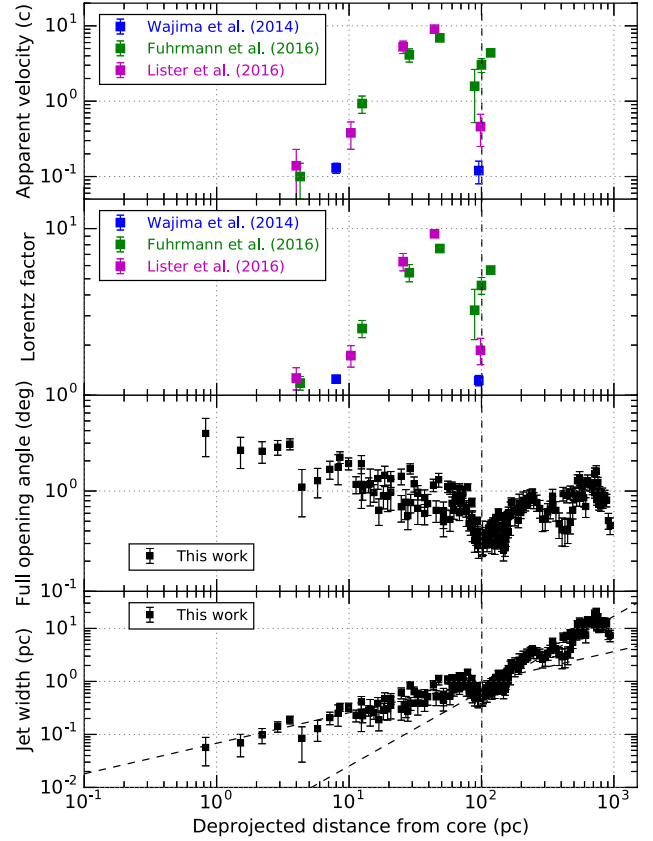}
 \caption{Collimation and acceleration of 1H\,0323+342 jet as a function of deprojected distance from the core. A jet viewing angle of 5\,deg is assumed. From the top panel to bottom: apparent motion reported in the literature; Lorentz factor calculated based on the apparent motion; intrinsic full opening angle calculated from the observed jet width profile; and jet width profile.}
 \label{fig:acc_col}
\end{figure}

In Figure\,\ref{fig:acc_col}, we show a combined plot of the proper motions in the literature, Lorentz factors ($\Gamma$) calculated based on the proper motion results, jet collimation profile and the corresponding full opening angle profiles, as a function of deprojected distance from the core. Here we adopt $\theta_{\rm view}=5$\,deg to calculate the deprojected distance. We also assume that the proper motions traced by jet components reflect bulk flow motions in the jet. 

We can indeed see an intriguing signature: the jet kinematics and global jet shape looks closely related with each other. The jet speed progressively increases with distance from a few to $\lesssim$100\,pc, where the jet shape is parabolic and the opening angle smoothly decreases with distance. Then the jet speed becomes the highest (at $\Gamma\sim$10) right before reaching $\sim$100\,pc where the stationary feature S is located. Beyond this distance, the jet speed seems to be saturated and the jet expands conically. Following M87~\citep{nakamura2013, asada2014}, this is the second example where the coincidence of collimation and acceleration as well as a recollimation shock at the end of ACZ was discovered. 

According to the current theoretical paradigm of powerful AGN jet formation, a jet is initially produced by strong magnetic fields amplified by a spinning black hole or/and an inner part of the accretion disk~\citep{bz1977, bp1982}. The flow is initially slow in a magnetically-dominated state, and the subsequent acceleration is realized by gradually converting the magnetic (and also thermal) energy into the kinetic one, while the ambient pressure from the external medium collimates the flow. As a consequence, an acceleration and collimation zone is co-formed over long distances from the black hole (by either magnetically \citep{mckinney2006, komissarov2007, komissarov2009} or hydrodynamically \citep{gomez1997, perucho2007}).

To test whether the observed collimation and acceleration zone in 1H\,0323+342 is causally connected or not, it is important to examine if the product $\Gamma\phi/2$ satisfies the condition $\Gamma\phi/2 \lesssim \sigma_m^{1/2}$, where $\phi/2$ and $\sigma_m$ is half-opening angle and magnetization parameter of the jet, respectively. This condition can be derived by requiring that the opening angle of a magnetized jet must be smaller than the opening angle of the Mach cone of fast-magnetosonic waves so that the flows can communicate across the jet~\citep{komissarov2009}. If the jet far beyond the central engine is roughly in equipartition (i.e., $\sigma_m\sim 1$), the condition becomes $\Gamma\phi/2\lesssim 1$. We calculated the product for the five components of \citet{lister2016}, and found $\Gamma \phi/2$$\sim$(0.01, 0.02, 0.03, 0.04, 0.02) for components (6, 5, 4, 3, 2). This indicates that these are all well causally connected. It is also notable that the values are roughly constant from $\sim$4 to $\sim$100\,pc, suggesting that the same acceleration/collimation efficiency is maintained over these distances. 
 
We also note that statistical studies for large samples of blazars indicate $\Gamma\phi/2$$\sim$0.2--0.3~\citep{clausen2013, pushkarev2017, jorstad2017}, a factor of 10 larger than those we estimated in 1H\,0323+342. If $\Gamma\phi/2\sim0.2$ is the case for 1H\,0323+342, this implies that the true bulk flows may be much faster (up to $\Gamma\lesssim100$) than those suggested in the observed proper motions.

\subsection{$\gamma$-ray Production at the End of ACZ?}
1H\,0323+342 exhibits one of most active $\gamma$-ray flares among the known $\gamma$-ray detected NLS1s~\citep{dammando2016, paliya2018}, but the exact location of the flaring region is still under debate. In general active $\gamma$-ray emission from AGN jets and blazars are thought to be produced in compact regions near the central engine ($\lesssim$subpc scales) due to the observed rapid variability time scales~\citep[e.g.,][]{tavecchio2010}. This scenario seems to be favored also for NLS1s based on the modeling of radio-to-$\gamma$-ray SED~\citep[e.g,][]{abdo2009b,paliya2014,dammando2016}. As for 1H\,0323+342, \citet{angelakis2015} report that the source is more variable with higher amplitude at mm frequencies than at low frequencies. Indeed, our VLBA 43\,GHz image shows the mm-wave emission is dominated by the compact core, and our core-shift measurements suggest that the location of the 43\,GHz core is not farther than $\sim$1\,pc from the black hole. This may naturally prefer the scenario that the $\gamma$-ray emission from this source is associated with the mm-wave core near the black hole. 

Nevertheless, our discovery of the extended ACZ with a recollimation shock might propose an alternative scenario for the high-energy site in this source. As also discussed in \citet{doi2018}, the idea is similar to that proposed for HST-1 in M87. The presence of the extended ACZ means that the cross section of the jet remains quite small even at large distances from the black hole. The cross section of S (at $z$$\sim$100\,pc, deprojected) is still within subpc (0.4\,pc), with the corresponding intrinsic opening angle reaching minimum at this distance ($\phi$$\sim$0.3\,deg). This value of $\phi$ is much smaller than the averaged $\phi$ of non-$Fermi$-LAT-detected blazars \citep[$\phi_{\rm LAT}$$\sim$2.5\,deg;][]{pushkarev2017} and close to or possibly even smaller than the averaged $\phi$ of LAT-detected $\gamma$-ray blazars \citep[$\phi_{\rm LAT}$$\sim$1.2\,deg;][]{pushkarev2017}. In fact, \citet{doi2018} found that the  multiple strong $\gamma$-ray flaring events occurred in 2016~\citep{paliya2014} coincided with the passage of the fastest superluminal ($\sim$9\,$c$) component through the feature S. Interestingly, during this period this source showed an enhanced power-law X-ray component that was likely from a nonthermal jet, while the thermal emission from the coronal/disk was relatively low~\citep{paliya2014}. Therefore, one may not preclude the feature S as a potential site of $\gamma$-ray emission in 1H\,0323+342.  

Note that one serious concern for this scenario is that the variability time scale based on the causality argument results in $t_{\rm var}= R(1+z)/c\delta\sim30$\,days, which cannot explain the day-scale $\gamma$-ray variabilities detected in the previous flares~\citep{paliya2014}. This might be reconciled if the feature S contains more compact substructures like HST-1~\citep{cheung2007, giroletti2012} or turbulent cells~\citep{marscher2014}, or if the jet contains the faster (larger $\delta$) flows (as discussed in Section 4.2). 

The production of $\gamma$-rays at large distances from the black hole (beyond BLR/torus) is proposed for BL Lacertae~\citep{marscher2008}, PKS\,1510-089~\citep{marscher2010} and OJ\,287~\citep{agudo2011} as well as M87/HST-1. These studies all suggest that a recollimation shock formed at the end of (or beyond) ACZ is a plausible site of $\gamma$-ray emission. Some of these studies suggest that $\gamma$-ray emission is generated also at another site closer to the central engine~\citep[e.g.,][]{hada2014}. In this case, one may expect different patterns of MWL correlation from different $\gamma$-ray events. For 1H\,0323+342, \citet{paliya2014} reported that some X-ray events were correlated with $\gamma$-ray while some other X-ray flares had no counterpart in $\gamma$-rays, suggesting the existence of multiple high-energy emission sites. A more dedicated VLBI monitoring of this jet and S along with high-energy instruments is the key to pin down the site(s) of $\gamma$-ray production definitively.

\subsection{Jet Collimation, Confinement Medium and Black Hole Mass}
The remarkable similarity of apparent jet structure between 1H\,0323+342 and M87 is surprising given the different $M_{\rm BH}$ estimation and different host galaxy type. However, a striking difference occurs when the spatial scale is measured in $R_{\rm s}$ unit instead of pc. Adopting $M_{\rm BH} = 2\times 10^{7}\,M_{\odot}$ and $\theta_{\rm view}=4-12$\,deg~\citep{fuhrmann2016, lister2016} for 1H\,0323+342, the deprojected distance of the jet transition (i.e., the feature S) from the nucleus results in (2--9)$\times 10^{7}\,R_{\rm s}$. This is tens to $\sim$100 times more distant than that of HST-1/M87. To maintain a collimated parabola jet shape, the pressure support from the external medium is essential~\citep[e.g.,][]{komissarov2009}. For M87 (and also NGC\,6251 and NGC\,4261), the structural transition occurs around $z$$\sim$$10^{5-6}\,R_{\rm s}$, which is near the Bondi radius or the sphere of gravitational influence (SGI) of the SMBH~\citep{asada2012, tseng2016, nakahara2018}. This suggests that the external gas bound by the SMBH potential supports the collimation of these jets. 

\begin{figure}[ttt]
 \centering
 \includegraphics[angle=0,width=1.0\columnwidth]{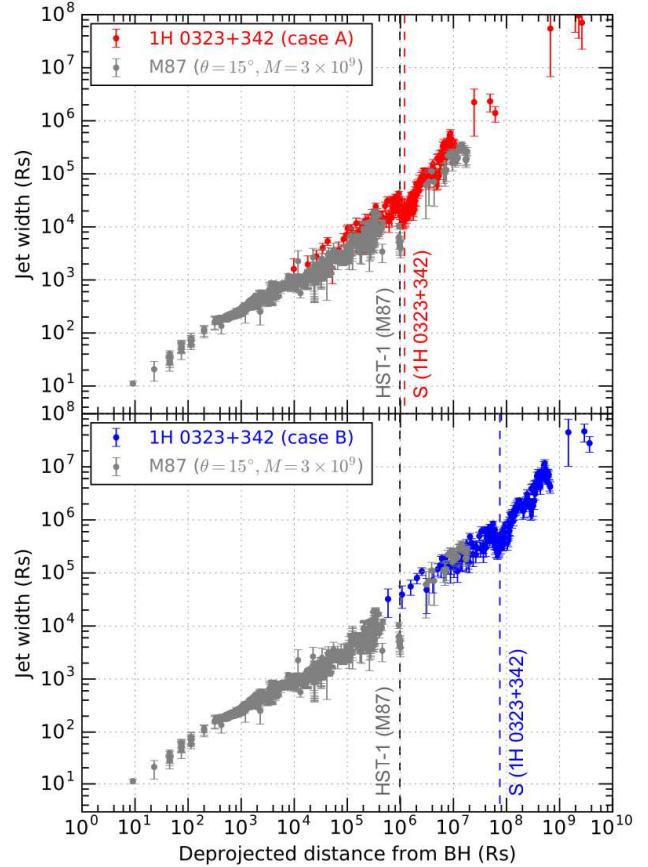}
 \caption{Comparison of jet collimation profile between 1H\,0323+342 and M87 in the unit of Schwarzschild radius. For 1H\,0323+342, we plot two different cases: (A) the red one for $(M_{\rm BH}, \theta_{\rm view})=(4\times 10^8\,M_{\odot}, 12\,{\rm deg})$ and (B) the blue one for $(M_{\rm BH}, \theta_{\rm view})=(2\times 10^7\,M_{\odot}, 3\,{\rm deg})$. The grey color data are for M87, which we combined the results from \citet{asada2012} and \citet{hada2013}. For M87 $(M_{\rm BH}, \theta_{\rm view})=(3\times 10^9\,M_{\odot}, 15\,{\rm deg})$ are assumed in this plot. The red and blue vertical dashed lines indicate the distance of S for the case (A) and (B), respectively. The grey vertical dashed line indicates the distance of HST-1.}
 \label{fig:m87vs1h}
\end{figure}

The discrepancy between 1H\,0323+342 and M87 can be reduced if the true $M_{\rm BH}$ of 1H\,0323+342 is larger than assumed. Indeed, the exact $M_{\rm BH}$ of this source is still highly controversial: while a small BH mass (a few $\times 10^{7}\,M_{\odot}$) seems to be generally preferred thanks to the consistency of various different independent methods~\citep{zhao2007, yao2015, wang2016, landt2017}, there are counterarguments suggesting a factor of 10--30 larger $M_{\rm BH}$, which is typical in blazars~\citep{tavares2014}. A similar discrepancy of $M_{\rm BH}$ estimation has been observed also for other $\gamma$-ray emitting NLS1s FBQS J1644+2619~\citep{dammando2017} and PKS\,2004-447~\citep{baldi2016}. In Figure~\ref{fig:m87vs1h}, we show a plot comparing the jet collimation profiles between M87 and 1H\,0323+342 for different $M_{\rm BH}$ and $\theta_{\rm view}$. Interestingly, if we assume ($M_{\rm BH}$, $\theta_{\rm view}$) = ($4\times 10^8\,M_{\odot}$, 12\,deg) for 1H\,0323+342 and ($M_{\rm BH}$, $\theta_{\rm view}$) = ($3\times 10^9\,M_{\odot}$, 15\,deg) for M87, the jet profiles of the two sources are remarkably overlapped with each other in $R_{\rm s}$ unit. This implies a common jet collimation mechanism is at work \textit{if} $M_{\rm BH}$ of 1H\,0323+342 is large. In this case, 1H\,0323+342 actually falls onto the famous $M_{\rm BH}-\sigma$ relation~\citep[$M_{\rm BH} \propto \sigma^{\sim4}$;][]{kormendy2013} along with M87. This implies that the SGI, defined by $r_{\rm SGI} = GM_{\rm BH}/\sigma^2$, could be expressed by $r_{\rm SGI}\propto M^{\sim1/2}_{\rm BH}$. Therefore the collimation of AGN jets from $M_{\rm BH}=10^{8-9}\,M_{\odot}$ might commonly break at distances around $r_{\rm SGI}\sim 3\times 10^5\,R_{\rm s}$. Note that the jet profile of 1H\,0323+342 shown by red color in Figure\,\ref{fig:m87vs1h} is also sufficiently overlapped with those of NGC\,6251~\citep{tseng2016, nakamura2018} and 3C\,273~\citep{akiyama2018}.

On the other hand, if the small $M_{\rm BH}$ is correct, the observed jet collimation would require additional/alternative external support that confines the jet until far beyond the SMBH potential. A likely source of such support would be large-scale winds/outflows surrounding the radio jet. Based on recent radiation-MHD simulations there are strong indications that super-Eddington BH accretion flows drive powerful outflows due to strong radiation pressure force from the radiatively very efficient accretion flows~\citep{ohsuga2009,ohsuga2011}. Recent systematic studies of [\ion{O}{3}] emission lines for samples of radio-loud NLS1s have found evidence of powerful outflows~\citep{berton2016a, komossa2016}. These outflows are associated with the narrow line region (NLR), whose spacial scales are supposed to be of the order of 1--100\,pc, consistent with the observed collimation scales of 1H\,0323+342. Therefore, if the small $M_{\rm BH}$ with high $\dot{M}$ is true for 1H\,0323+342, the observed 1--100\,pc-scale collimation might be supported by the NLR outflows. Note that the observed NLR outflows in radio-loud NLS1s might also be a consequence of the propagation of relativistic jets rather than driven by the accretion flows~\citep{berton2016a}. Future VLBI polarimetric observations with RM measurements would provide more insights into the jet-surrounding interactions in this source~\citep[e.g., as performed in 3C\,120;][]{gomez2008, gomez2011}.

Therefore in any case, our results on 1H\,0323+342 have very intriguing implications for jet collimation, black hole mass, accretion state, confinement medium/environment and their possible relationship with each other: (1) If the large $M_{\rm BH}$ expected from the $M_{\rm BH}-\sigma$ relation is the case, the jet collimation may be regulated by the gas bound by the SMBH potential, just as suggested for M87 and nearby radio galaxies. In contrast, (2) if the small $M_{\rm BH}$ (and thus the higher Eddington ratio) is the case, the jet collimation may be regulated by NLR outflows that are driven by super-Eddington BH accretion flows. Solving this question will allow us to build a unified picture of AGN jet formation mechanisms as a function of $M_{\rm BH}$ and $\dot{M}$, and also might help better understand the co-evolution and feedback between SMBH and host galaxies.

\section{Summary}

We reported results from a detailed nine-frequency VLBA observation of the jet in the nearest $\gamma$-ray detected radio-loud NLS1 galaxy 1H\,0323+342. We summarize our main results as follows.

\begin{enumerate}
\item We obtained multi-frequency VLBI images of the 1H\,0323+342 jet at various spatial scales from $\sim$100\,mas (at 1.4\,GHz) down to $\sim$0.1\,mas (at 43\,GHz). At high frequencies ($\gtrsim$24\,GHz) the radio core at the jet base dominates the emission, while at low frequencies ($\lesssim$2\,GHz) the quasi-stationary, rebrightened feature S at 7\,mas from the core becomes the brightest component. We found that S remained there over more than 10 years. 
\item We extensively investigated the width profile $W(z)$ of this jet over a wide range of distance from $z$$\sim$0.1\,mas to $z$$\sim$$10^4$\,mas using our VLBA images and supplementary MERLIN/VLA archival images. We discovered that the jet maintained a parabolic shape from the jet base to $z$$\sim$7\,mas where S is located, providing evidence of a jet collimation zone. Beyond $z$$\sim$7\,mas, the jet started to expand more rapidly, transitioning into a hyperbolic/conical shape.  
\item We found that the jet cross section of S (i.e., the site of jet transition) is significantly smaller (by a factor of 2) than that expected from the parabolic profile of the inner jet. Between the inner parabolic zone and the compressed S, we detected a ``recollimation zone'' connecting these two regions. Moreover, we found that S showed strongly polarized signals (as high as $\sim$30\,\%), possible spectral flattening, and passing of highly superluminal components. A set of these results suggest that the quasi-stationary feature S is a recollimation shock at the end of the collimation zone, producing highly ordered $B$-fields and re-energizing jet particles. 

\item By combining the collimation profile and the jet proper motions results in the literature, we discovered the coincidence of jet collimation and acceleration between the jet base and S. The product of Lorentz factor and (half) opening angle of this jet satisfies $\Gamma\phi/2<1$, indicating that the observed collimation and acceleration zone (ACZ) is causally connected.

\item Our discovery on ACZ showed that the intrinsic opening angle of the jet becomes minimum at S, with the Lorentz factor reaching maximum, suggesting a highly Doppler boosting jet. Together with the fact that an active $\gamma$-ray flaring event in 1H\,0323+342 coincided with the passage of a highly superluminal component through S~\citep{doi2018}, the recollimation shock S at the end of ACZ might be a potential site of the $\gamma$-ray production as well as the jet base. The deprojected distance of this feature is $\sim$40-100\,pc. The $\gamma$-ray production at such distances is much larger than those usually suggested in typical $\gamma$-ray blazars, while this situation is quite similar to HST-1 in the jet of M87. 

\item We found that the core-shift of the 1H\,0323+342 jet is small between 1.4 and 43\,GHz. This is consistent with the jet having a small viewing angle to our line-of-sight. We estimated a relative separation between the 43\,GHz core and the central BH by using the measured parabolic collimation profile of the inner jet, and obtained a deprojected separation of $\sim$(0.24--0.71)\,pc (or possibly shorter) for a viewing angle of 4--12\,deg. 

\item We found various remarkable similarities in the pc-scale jet structures between 1H\,0323+342 and M87 (e.g., the collimation profile and recollimation shock). If $M_{\rm BH}$ of 1H\,0323+342 is $\sim$$4\times10^8\,M_{\odot}$ and thus follows the $M_{\rm BH}-\sigma$ relation, we suggest that a common jet formation mechanism is at work in M87 and 1H\,0323+342, where the external gas bound by the SMBH potential plays a role in collimating the jets. 

\item If $M_{\rm BH}$ of 1H\,0323+342 is a few times of $10^7\,M_{\odot}$ (and thus $\dot{M}$ close to or above the Eddington limit), which is proposed by several previous papers, the parabolic jet continues up to $10^{7-8}\,R_{\rm s}$ from the black hole, suggesting additional large-scale structures that confine the jet. NLR outflows, which may be driven by super-Eddington accretion flows,  could be responsible for the large-scale jet confinement.  

\end{enumerate}

\acknowledgments 

We sincerely thank the anonymous referee for his/her careful reviewing that improved the manuscript. The Very Long Baseline Array is an instrument of the Long Baseline Observatory. The Long Baseline Observatory is a facility of the National Science Foundation operated by Associated Universities, Inc. This work made use of the Swinburne University of Technology software correlator~\citep{deller2011}, developed as part of the Australian Major National Research Facilities Programme and operated under license. This research has made use of data from the MOJAVE database that is maintained by the MOJAVE team~\citep{lister2009}. Part of this work was done with the contribution of the Italian Ministry of Foreign Affairs and University and Research for the collaboration project between Italy and Japan.

\section*{Appendix}
\section*{Calibrator 0329+3510}

\begin{figure*}[ttt]
 \centering
 \includegraphics[angle=0,width=1.0\textwidth]{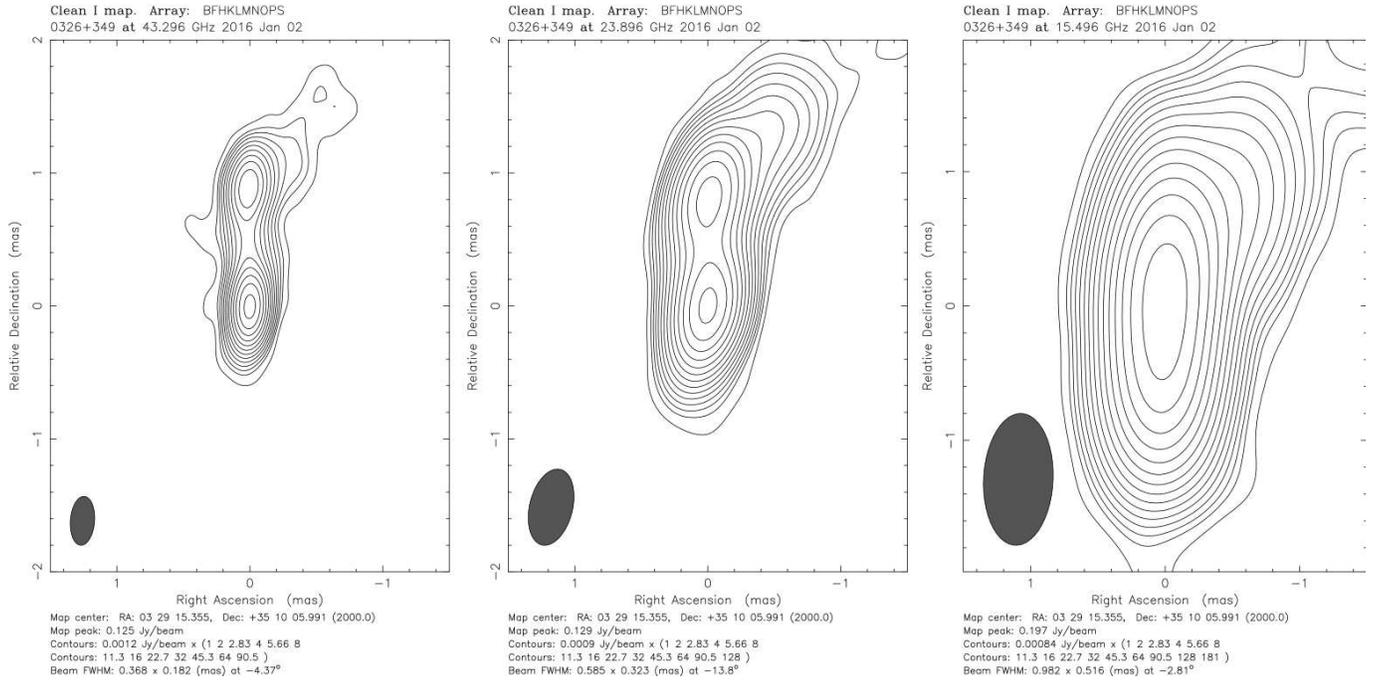}
 \caption{Self-calibrated VLBA images of 0329+3510. From the left to right, the images are at 43, 24 and 15\,GHz, respectively. For each panel, a restoring beam is shown at the bottom left corner of the map, and contours start from $-1$, 1, 2,...times a 3$\sigma$ image rms level and increase by factors of $2^{1/2}$.}
 \label{fig:0326}
\end{figure*}

0329+3510 is a radio source separated by 1.36\,deg on the sky from 1H\,0323+342.  This source is among candidates of High Frequency Peakers which are a subclass of young radio galaxies~~\citep{dallacasa2000}. In Figure~\ref{fig:0326} we show self-calibrated images of 0329+3510 obtained at 15, 24 and 43\,GHz. The source morphology significantly changes with frequency. At 43\,GHz where the highest angular resolution is available, the source is resolved into the core (the southernmost brightest region) and a strong knot. The core and the knot is also securely resolved at 24\,GHz. However, the two features are mixed in a synthesized beam at 15\,GHz and the lower frequencies. This is because the knot becomes brighter as frequency decreases (i.e., an optically-thin steep spectrum) while the core spectrum is relatively flat or slightly inverted (i.e., optically-thick), as well as the resolution effect. This prevents us from specifying a reliable reference position in the 0329+3510 images at $\leq$15\,GHz. 

Therefore, we employed relative astrometry between 1H\,0323+342 and 0329+3510 only at 24 and 43\,GHz. At each frequency, we measured the core position of 1H\,0323+342 with respect to the core of 0329+3510. In this case, we need to know the effect of core-shift of the calibrator itself since the observed core-shift on the target image is the mixture of the core-shift from both sources.  Actually, 0329+3510 has a well-defined optically-thin knot at 24/43\,GHz that should be stationary with frequency, so we can measure the intrinsic core-shift of 0329+3510 based on the self-referencing method using the self-calibrated images. As a result, we found that the intrinsic core-shift of 0329+3510 between 24 and 43\,GHz is virtually negligible (less than 6\,$\mu$as). Therefore, the observed core-shift of 1H\,0323+342 with respect to the core of 0329+3510 is virtually represents the intrinsic core-shift of the target.

\end{document}